\documentclass[12pt]{article}
\usepackage{mathtext}
\usepackage{graphicx}
\usepackage[T2A]{fontenc}
\usepackage[utf8]{inputenc}
\usepackage{amsfonts}
\usepackage{amssymb}
\usepackage{amsmath}
\usepackage{pgfplots}
\usepackage{braket}
\usepackage{tikz}
\usepackage{float}
\usepackage{subcaption}
\begin{document}
\title{Remote Stimulation of QED Scenarios in the Jaynes-Cummings-Hubbard Model}

\author{Andrey Kuzminskiy (1),\\ Yuri Ozhigov (1)\\
{\it 
1. Lomonosov Moscow State University,
} \\
{\it Faculty of Computational Mathematics and Cybernetics,}
\\
}
\maketitle

\begin{abstract}
The article addresses the important and relevant task of remote induction of quantum dynamic scenarios. This involves transferring such scenarios from donor atoms to a target atom. This induction is based on the enhancement of quantum transitions in the presence of multiple photons of the same transition. We use the quantum master equation for the Tavis-Cummings-Hubbard (TCH) model with multiple cavities connected to the target cavity via waveguides. The dependence of the efficiency and transfer of the scenario on the number of donor cavities, the number of atoms in them, and the bandwidth of the waveguides is investigated.

\end{abstract}
\newpage

\section{Introduction}
Remote transfer of genetic information is one of the most intriguing dynamic scenarios in the microcosm. Experiments of this kind have been conducted in several laboratories (see, for example, the work \cite{mont} by the team led by Luc Montagnier), which demonstrated not only the presence of the effect but also the non-trivial conditions required for its manifestation. The latter has sparked a wave of criticism from the scientific community (e.g., the work \cite{scept}). This highlights the importance of both clarifying the specific conditions for remote transfer and understanding the mechanism behind this phenomenon.

It is unlikely that an explanation for remote transfer can be found without invoking quantum mechanics. Particularly, the interaction of individual photons with atoms which is most conveniently described within the framework of the Tavis-Cummings-Hubbard (TCH) model (\cite{TC}, \cite{T}). This model describes the interaction of a multimode field with atoms distributed across optical cavities connected by waveguides.

The mechanism we propose for explaining remote transfer is based on the significant enhancement of the transition energy between atomic states in the presence of a large number of photons of the same transition near the atom. We refer to this as transition induction and the transition itself as induced.

\section{Remote State Transfer}

Remote transfer of quantum states with the movement of only classical information is called teleportation. Since the pioneering works \cite{Tel0}, \cite{Tel} teleportation has attracted constant interest, both in connection with cryptography (see, for example, \cite{Tel1}) and in its own right as a method of information transfer using quantum communication channels. This is particularly relevant for organizing quantum networks (\cite{netw}) and distributed computing (\cite{gates}, \cite{transfer}). There are also numerous works dedicated to specific physical systems where the influence of remote state transfer is important: \cite{water}, \cite{mont}, particularly in the biological sphere—\cite{dna}, \cite{decoh}, \cite{recogn}.

We will consider a simpler method of creating a unidirectional character based on the Jaynes-Cummings-Hubbard Hamiltonian. This Hamiltonian describes the dynamics of quantum states of ensembles of two-level atoms distributed across optical cavities connected by waveguides. Our model allows for the representation of ensemble dynamics without optical cavities where photons are removed from one location and can enter another. This movement of photons under certain conditions can create an interesting effect of remote stimulation of quantum electrodynamics (QED) scenarios which we will demonstrate using our model.

In computer modeling we used the quantum master equation which demonstrates the high accuracy of our calculations. These calculations were performed using the MSU-270 supercomputer.

\section{Simplest Example of Photonic Transfer Dynamics}

Remote transfer of quantum states (see, for example, \cite{transfer}) is a crucial effect whose role is not yet fully understood. We will consider a related effect of remote stimulation of QED scenarios and illustrate the effectiveness of a purely unitary model using this example.

\begin{figure}[H]
\centering
\includegraphics[scale=1.3]{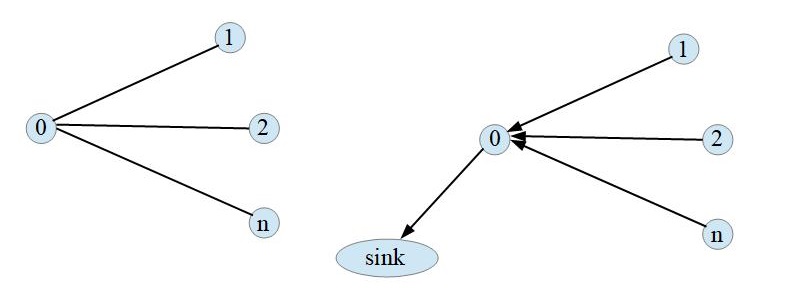}
\caption{General scheme of remote induction of a dynamic scenario}
\label{fig:f1a }
\end{figure}

Consider the simplest example of a transition between levels in one cavity, induced by photons coming from similar systems in the environment. The basis will be the configuration shown in Figure \ref{fig:f1a }, where we consider the transfer of photons from cavities $1, 2, … , n$ to cavity $0$. The initial state is chosen as $|0,1,1,...,1\rangle$, where the cavity numbering starts from zero. Unitary oscillation under the constant Jaynes-Cummings-Hubbard Hamiltonian for the system of cavities on the left side of Fig. \ref{fig:f1a } leads to the alternation of states $|0,1,1,...,1\rangle\leftrightarrow |n,0,0,...,0\rangle$. Thus, photons leaving cavities $1,2,...,n$ all end up in cavity $0$. If we place enough atoms in the cavities such that the transition induced by a photon of a given mode occurs much slower in cavity $0$ than in the other cavities, this transition can become more probable due to the presence of many photons in cavity $0$.

We divide the transition $|0,1,1,...,1\rangle\leftrightarrow |n,0,0,...,0\rangle$ into two segments of equal duration. The first transition $|0,1,1,...,1\rangle\rightarrow |n,0,0,...,0\rangle$ corresponds to filling cavity zero and the second transition 
$|0,1,1,...,1\rangle\leftarrow |n,0,0,...,0\rangle$ corresponds to the outflow of photons into the sink cavity. The latter process can be identified with the emission of photons from cavity 0 into the surrounding space. Scenario concludes on this. The presence of photons of the resonator mode starting from 1 enhances the corresponding transition so that it begins to dominate over the intrinsic transition in cavity zero—\ref{fig:f2a } below. This is the remote transfer of dynamics.

It is also possible to have several three-level systems in the target location as shown at the bottom of Figure \ref{fig:f2a }. Finally, we can consider the remote induction of dynamic scenarios in multilevel systems. This effect is applicable. For example in describing the process of ice crystal formation. Here each level corresponds to a specific partial crystal so that in a large mass of water molecules remote induction of dynamic scenarios of a very specific type will occur. This will lead to uniform ice crystals.

\begin{figure}[H]
\centering
\includegraphics[scale=1.3]{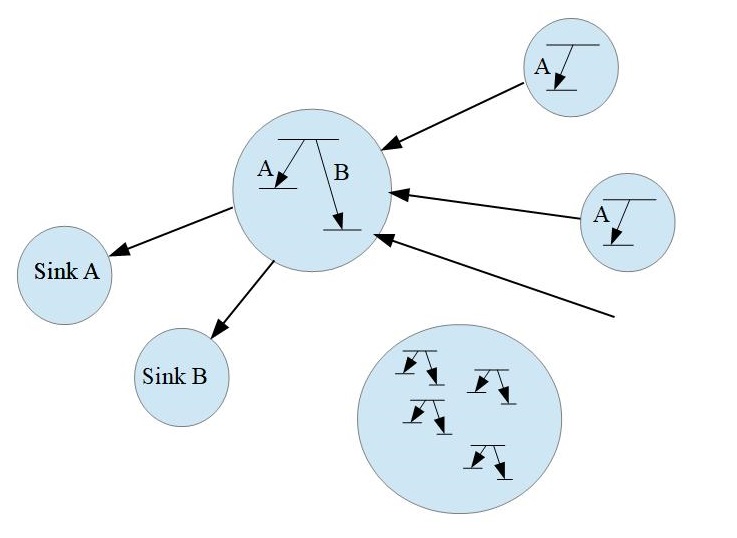}
\caption{Transfer of a dynamic scenario with suppression of the basic transition to level B}
\label{fig:f2a }
\end{figure}

The numerical characteristic of remote stimulation is the probability of photons entering sink A compared to sink B. The filling of sink B predominates if there are no cavities on the right (which can be called stimulating cavities). The presence of stimulating cavities creates an advantage in filling sink A. We called this effect remote stimulation.

\section{Simulation Results}

Several configurations of systems with three-level atoms were modeled.

The first configuration consisted of cavity 0 with n atoms in the highest energy states and cavity 1 with m photons with transition A from Figure \ref{fig:f2a }. The results showed only a slight deviation from the original scenario without cavity 1 with photons. With the emission of photons, there is no final change. The reason for this result is that photons with the mode of transition A enhance the reverse transition to the highest energy state which has the most probable transition B.

\begin{figure}[H]
\centering
\begin{subfigure}[t]{0.45\textwidth}
\includegraphics[width=\textwidth]{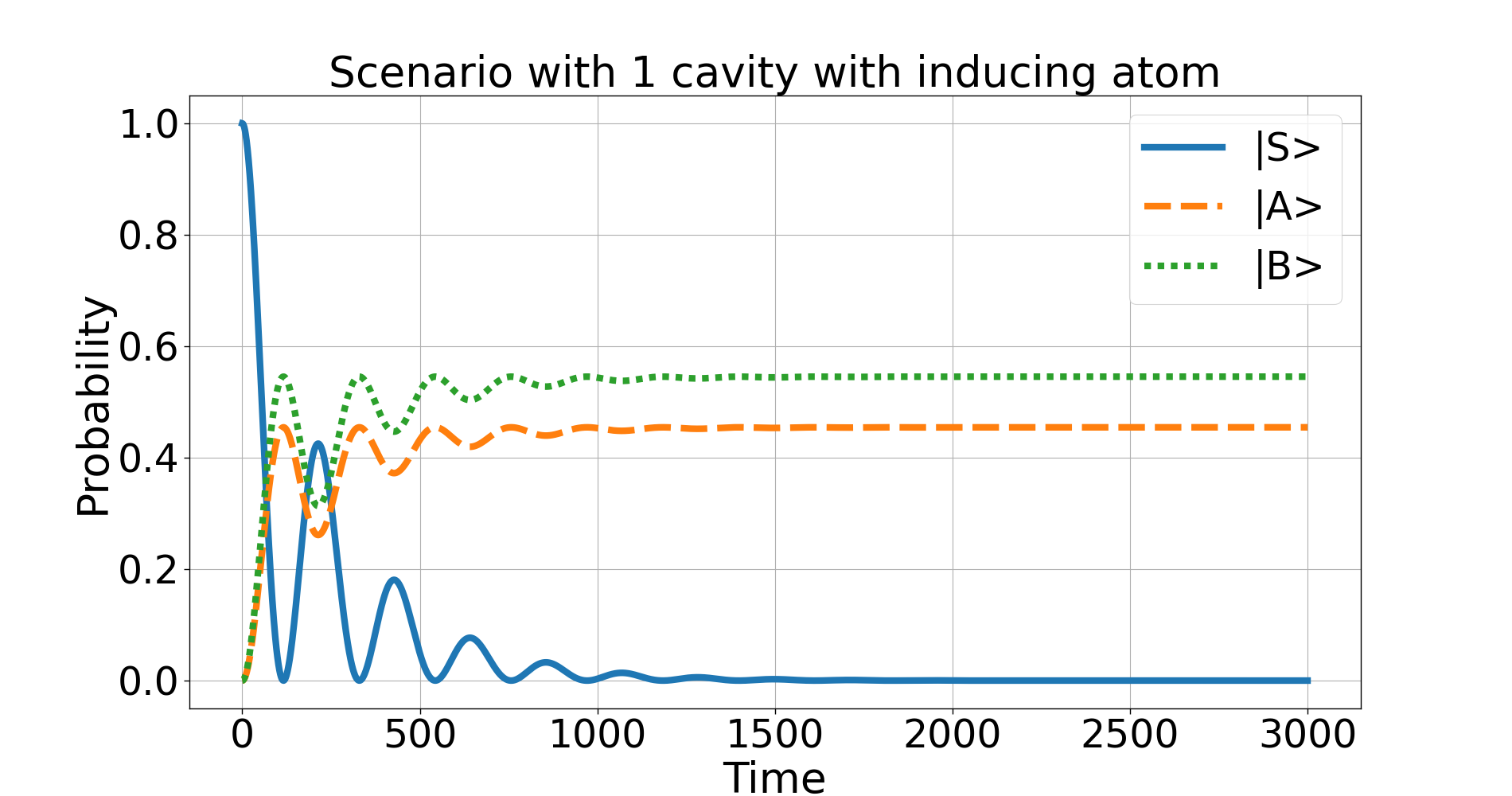}
\caption{Original scenario with 2 three-level atoms in cavity 0 in the initial highest energy states $|S\rangle$}
\label{fig:orig }
\end{subfigure}
\hfill
\begin{subfigure}[t]{0.45\textwidth}
\includegraphics[scale=0.4, width=\textwidth]{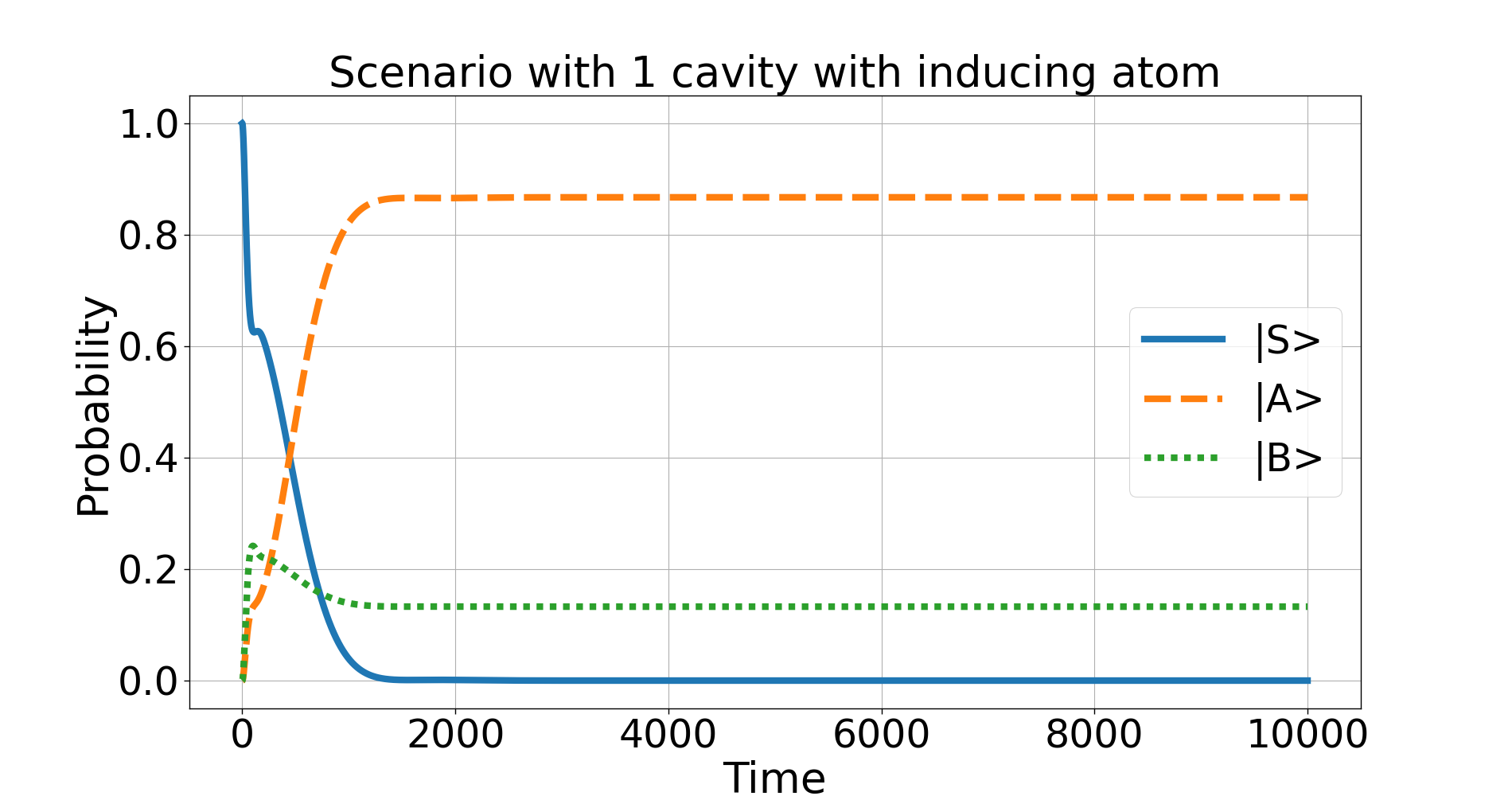}
\caption{Addition to the original scenario of a cavity with atoms with only allowed transitions A.}
\label{fig:distant }
\end{subfigure}
\caption{Demonstration of the effect on three-level atoms}
\end{figure}

The second configuration is successful. Cavity 0 with n atoms in the highest energy states with all possible transitions and cavity 1 with $m$ atoms in the highest energy states with only transition A possible. In this configuration the effect of remote stimulation is clearly manifested. Below are the results with a single number of atoms per cavity (increasing their number does not affect the final result except for slowing down the growth of the probability of the desired state $|A\rangle$. The limits themselves do not change). On Figure \ref{fig:orig } only 0 cavity is. On Figure \ref{fig:distant } the result with the effect of remote stimulation is.

\begin{figure}[H]
\centering
\begin{subfigure}[t]{0.45\textwidth}
\includegraphics[width=\textwidth]{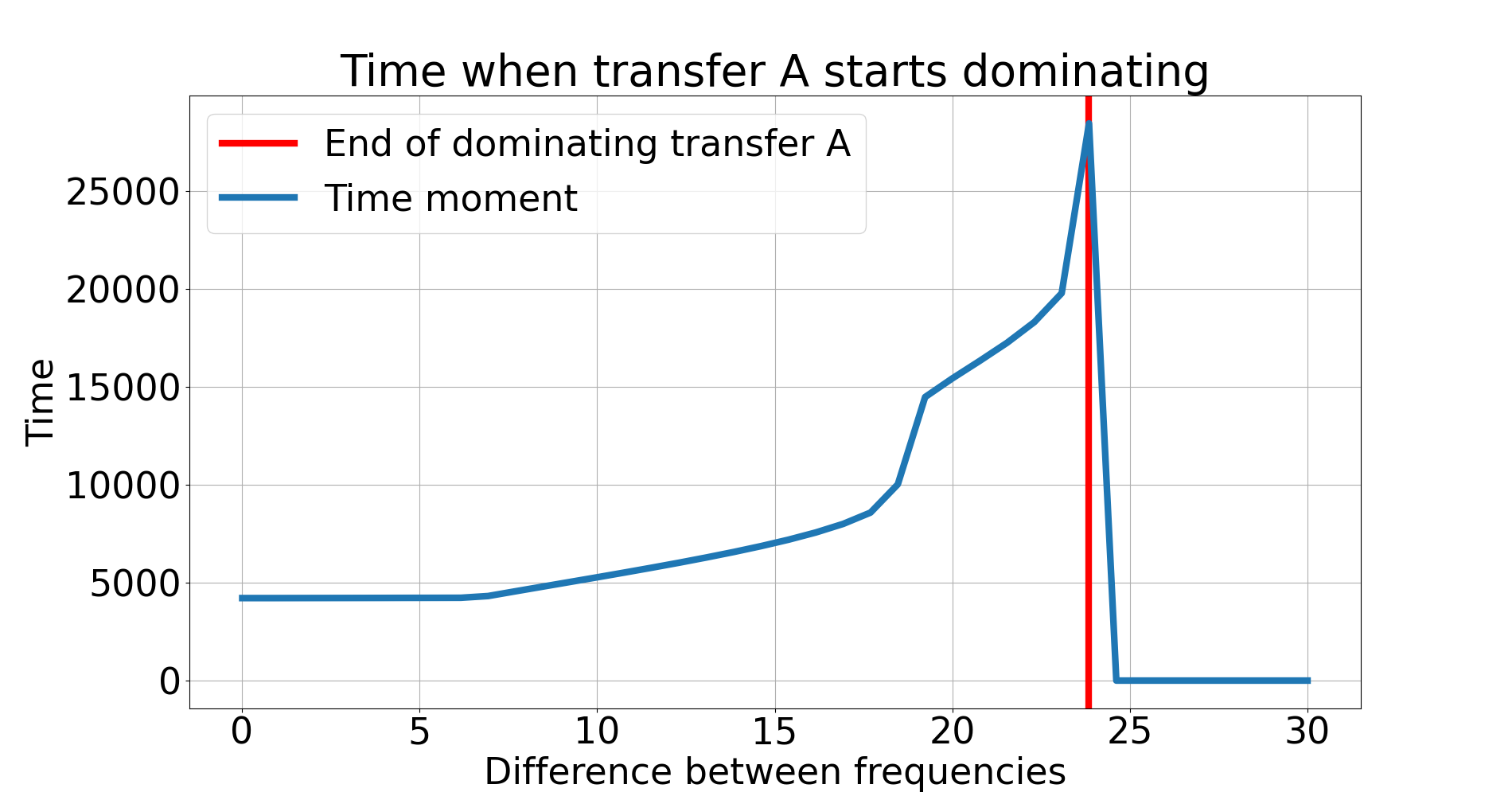}
\caption{Graph of the times when the probability of transition A of an atom in cavity 0 becomes greater than all other outcomes, i.e., states $|S\rangle$ and $|B\rangle$, depending on the frequency differences between these levels}
\label{fig:en_time }
\end{subfigure}
\hfill
\begin{subfigure}[t]{0.45\textwidth}
\includegraphics[width=\textwidth]{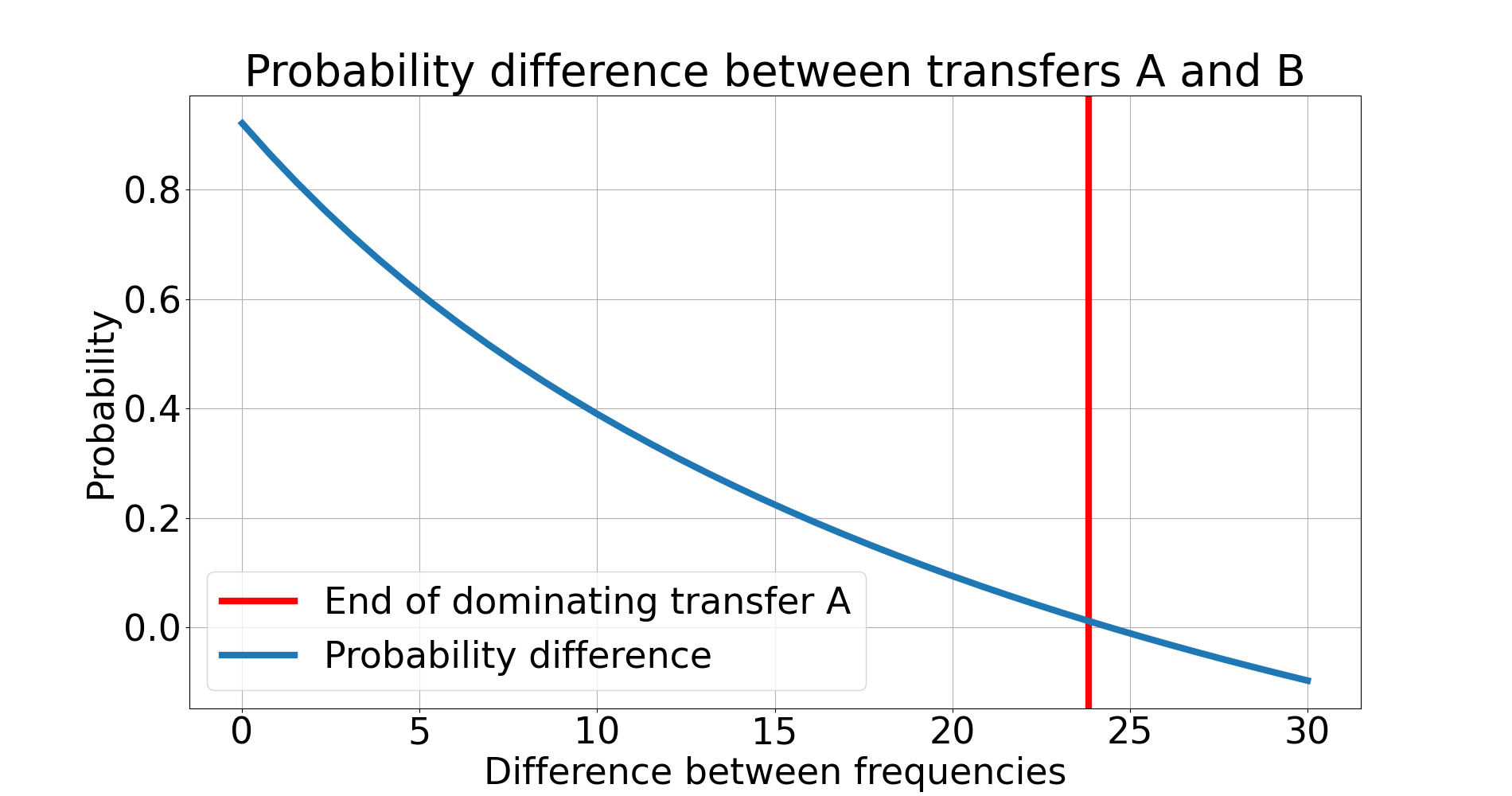}
\caption{Difference in probabilities between states 
$|A\rangle$ and $|B\rangle$}
\label{fig:en_prob_diff }
\end{subfigure}
\caption{Graphs of the effect's dependence on frequency differences}
\end{figure}

Now the dependencies of this effect on various parameters are demonstrated. Figure \ref{fig:en_time } shows the time when the probability of transition A becomes greater than all other outcomes. Figure \ref{fig:en_prob_diff } shows the difference in probabilities as they approach infinity (i.e., asymptotes) between transitions A and B. They show the nature of this effect depending on the frequency difference between levels A and B. It can be seen that this effect is quite stable over a wide range of energy differences. In this configuration it ceases to work when the energy of transition to level B exceeds that of transition to level A by 23 times.

\begin{figure}[H]
\centering
\begin{subfigure}[t]{0.45\textwidth}
\includegraphics[width=\textwidth]{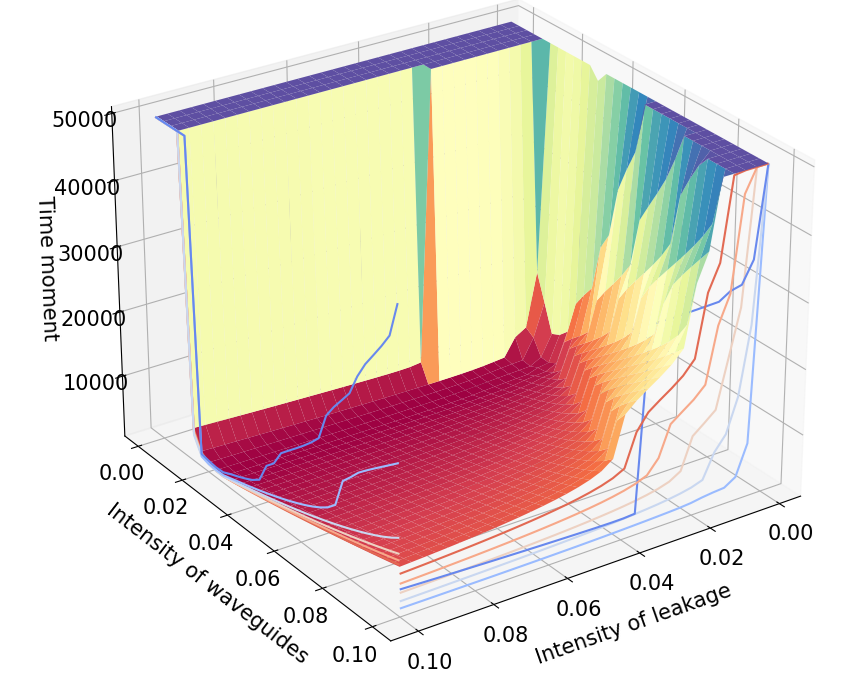}
\caption{Graph of the times when the probability of transition A of an atom in cavity 0 becomes greater than all other outcomes, i.e., states $|S\rangle$ and $|B\rangle$, depending on the waveguide and photon leakage intensities}
\label{fig:w_time }
\end{subfigure}
\hfill
\begin{subfigure}[t]{0.45\textwidth}
\includegraphics[width=\textwidth]{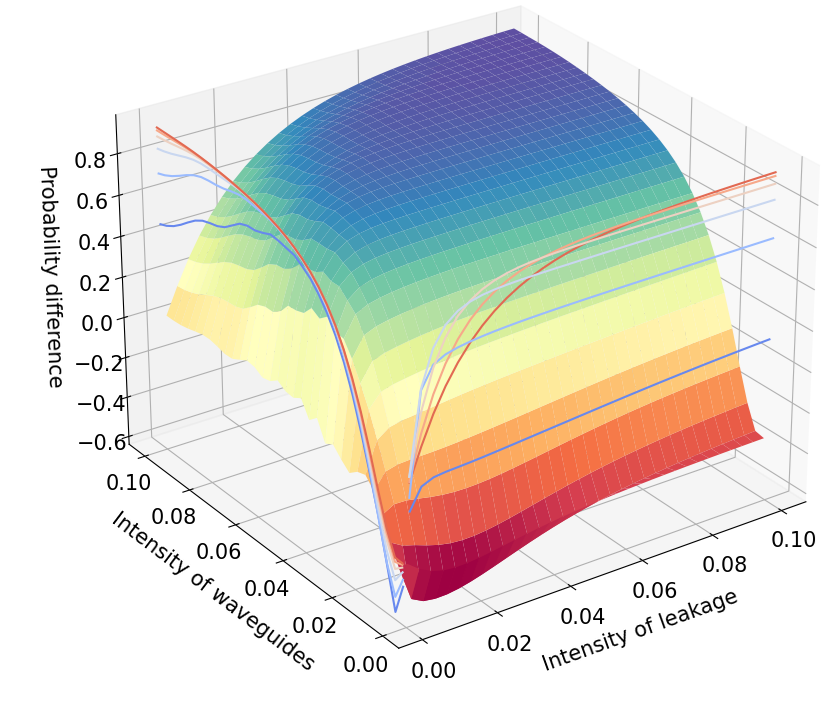}
\caption{Difference in probabilities between states 
$|A\rangle$ and $|B\rangle$}
\label{fig:w_prob_diff }
\end{subfigure}
\caption{Graphs of the effect's dependence on waveguide and photon leakage intensities}
\end{figure}

Now consider the dependence on the intensities of the waveguide between cavities 0 and 1, as well as on the intensity of photon leakage from cavity 0. Figure \ref{fig:w_time } shows the same times as in Figure \ref{fig:en_time }. It is clearly seen that the waveguide intensity accelerates the effect only up to a certain limit, much like in the quantum bottleneck effect. The results also showed that increasing the intensity of photon leakage from cavity 0 not only enhances our effect but also makes the entire dynamics more stable (i.e reduces the number of oscillations in it).

Figure \ref{fig:w_prob_diff } shows the difference in probabilities between transitions A and B at infinity. Here, the negative difference is also indicated, i.e., the probability of transition A does not become greater than B (or becomes so after a very long time). It can be seen that increasing both parameters increases the difference and, accordingly, enhances our effect. If we compare the difference in probabilities with the time, we can see that the most optimal option in terms of the speed of the probability shift does not lead to the largest difference in probabilities, i.e., the probability increases with increasing both parameters, but the speed of achieving this difference slows down.

The source code is available in the repository \cite{stimul_code}, written using the QComputations library \cite{QComputations}.

The complexity of the calculations performed largely lies in solving the quantum master equation. The Runge-Kutta method of the 2nd order was used for the solution parallelized at the level of block-distributed matrices. The Hamiltonian was generated using the selection method, which generated the basis. The size of the system grows approximately as the sum of all binomial coefficients with the increase in the number of photons and atoms, i.e., as $О(2^{p + m + c})$, where $p$ is the initial number of photons, $m$ is the total number of atoms and $c$ is the number of cavities. The complexity becomes approximately equal $О(\frac{2^{64(p + m + c)}}{P})$ taking into account the use of $P$ cores and the complexity of the Runge-Kutta method with increasing system growth. The acceleration is taken from the description of the SUMMA algorithm for matrix multiplication \cite{SUMMA}. It is used in the Intel MKL library, on which the QComputations library \cite{QComputations} is based.

\section{Conclusions}

The simulation results demonstrate the significance of the remote stimulation effect despite significant differences in transition frequencies (more than 20 times). The article above presents the results and figures with one donor cavity with 1 inducing atom in it and with 1 atom in the 0 cavity. In general, the model was tested on a system of up to 3 cavities with 1 atom each, as well as up to 2 atoms with 1 cavity. They show that the presence of photons with lower energy in the necessary cavities can significantly increase the probability of a less probable initial scenario. This effect was also investigated depending on the parameters of waveguide intensities and photon leakage. The results demonstrate that with not very large differences in transition energies we can fully control this effect. Also we can regulate the speed and final probability with adiabatic calculations. As a result, it turns out that the effect is strongest with a small difference in energies between the transition energies as well as with maximum values of waveguide intensities and decoherence (photon leakage).

\section{Acknowledgments}

The calculations in this work were performed on a personal computer with 16 threads (Intel i9-9900K). The role of the supercomputer is to verify the adequacy of the model. This means that the effect found on a low-performance machine is refined when working on a supercomputer. The adequacy of the TCH model used by us must be verified by calculations on a high-performance computer. It is a task for the near future.

We would like to express our deep gratitude to the administration of the MSU-270 supercomputer for providing the resources for this goals.

\end{document}